\documentclass{llncs}
\usepackage[utf8]{inputenc}
\usepackage[T1]{fontenc}
\usepackage{booktabs}
\usepackage{longtable}
\usepackage{array}
\usepackage{hyperref}
\usepackage{graphicx}
\usepackage{enumitem}
\usepackage{amsmath}
\usepackage{tikz}
\usetikzlibrary{positioning}
\usepackage[inkscapelatex=false]{svg}
\usepackage{microtype}
\usepackage{marvosym}
\title{CyBOKClaw: Human-in-the-Loop CyBOK Mapping for Cybersecurity Curriculum}
\author{Yan Lin Aung\textsuperscript{(\Letter)}\and
Kevin Togbe}
\institute{
University of Derby, Derby, UK\\
\email{y.aung@derby.ac.uk, k.togbe1@unimail.derby.ac.uk}
}
\begin{document}
\maketitle
\begin{abstract}
This paper presents \texttt{CyBOKClaw}, an interpretable human-in-the-loop retrieval framework for mapping cybersecurity keywords or phrases (KWoPs) to the Cyber Security Body of Knowledge (CyBOK).
Rather than treating the task as strict exact classification, the framework is designed as a top-$k$ candidate generator for expert review. It combines query normalization, curated term expansion, concept-level boosts, topic-description enrichment, and domain-sensitive ranking rules. 
Because educational KWoPs are often broad, ambiguous, and only approximately aligned with CyBOK terminology, strict exact matching provides only a partial account of practical utility. 
We therefore evaluate the framework using both structural retrieval metrics and an expert-guided top-5 usefulness metric, ECA-5 (Exact or Closest Acceptable Match at top-5), which records whether the returned candidates contain at least one mapping that an expert would judge exact or accept as the nearest practical CyBOK placement.
On the development dataset, \texttt{CyBOKClaw} achieves 64.73\% EXA-5 (Exact Match at top-5), 84.18\% structural semantic alignment, and 91.88\% ECA-5; on the validation dataset, it achieves 81.19\% EXA-5, 93.32\% structural semantic alignment, and 98.00\% ECA-5.
These results show that expert-guided top-$k$ usefulness provides a more faithful account of practical CyBOK mapping utility than exact structural matching alone, and that \texttt{CyBOKClaw} is effective as a CyBOK-specific expert-support retrieval system.
\end{abstract}
\section{Introduction}
The Cyber Security Body of Knowledge (CyBOK) provides a structured taxonomy for organizing cybersecurity knowledge through Knowledge Areas, Topics, and Indicative Material (IM)~\cite{cybok_official}.
As cybersecurity education expands across university curricula and professional training environments, there is a growing demand for methods that can align curriculum terms, lecture topics, learning outcomes, and assessment materials with CyBOK.
Such alignment can support curriculum design, programme review, syllabus comparison, educational knowledge organization and certification activities~\cite{harden2001mapping,cyber_education_framework}.

In practice, however, mapping natural-language cybersecurity keywords or phrases (KWoPs) to CyBOK is rarely a straightforward exact-classification task.
Many KWoPs are broad, ambiguous, lecture-title-like, or only approximately aligned with CyBOK terminology.
Under these conditions, the most useful system behaviour is often to return a strong set of candidate mappings for expert review rather than forcing a single exact mapping.
Strict tuple-level matching therefore provides only a partial account of practical system performance.

This paper presents \texttt{CyBOKClaw}, a human-in-the-loop retrieval framework for mapping cybersecurity KWoPs to CyBOK candidates.
The framework is designed as a top-$k$ mapping assistant and combines heuristic query normalization, curated term expansion, concept-level boosts, topic-description enrichment, and domain-sensitive ranking rules to surface semantically plausible CyBOK mappings for human review.
We study two related research questions: first, whether a lightweight curated retrieval framework can provide useful top-$k$ CyBOK candidates for curriculum mapping; and second, whether expert-guided usefulness metrics reveal practical value that strict tuple-level metrics miss.

Exact structural metrics alone do not adequately reflect practical usefulness in this setting.
EXA-5 (Exact Match at top-5) measures tuple-level retrieval, while structural semantic alignment captures algorithmic proximity to the reference mapping.
Neither directly captures whether the returned candidates would actually help an expert complete the mapping task.
To address this gap, we adopt an expert-guided evaluation centred on ECA-5 (Exact or Closest Acceptable Match at top-5), which records whether the returned top-5 contains at least one candidate an expert would judge as exact or accept as the nearest practical CyBOK placement.

Experimental results reveal a substantial gap between structural matching metrics and expert-guided usefulness.
On the development dataset, \texttt{CyBOKClaw} achieves 64.73\% EXA-5, 84.18\% structural semantic alignment, and 91.88\% ECA-5.
On the validation dataset, it achieves 81.19\% EXA-5, 93.32\% structural semantic alignment, and 98.00\% ECA-5.
These findings indicate that exact tuple-level retrieval alone understates the practical value in supporting expert-driven CyBOK mapping.

The primary contribution of this work is a CyBOK-specific formulation of curriculum mapping as a human-in-the-loop top-$k$ retrieval problem.
We develop a practical retrieval framework aligned with the CyBOK Knowledge Tree, introduce a combined structural and expert-guided evaluation methodology, and show empirically that exact-match metrics substantially understate expert-facing usefulness in this setting.
\section{Contributions}
The principal contributions of this work are as follows:
\begin{itemize}[noitemsep]
\item \textbf{A practical CyBOK mapping system.} We present CyBOKClaw, a semantic retrieval system that maps cybersecurity KWoPs to CyBOK Knowledge Area (KA), Topic, and Indicative Material (IM) candidates.

\item \textbf{A human‑in‑the‑loop formulation of CyBOK mapping.} We conceptualize CyBOK mapping as a top‑$k$ candidate‑generation task rather than a strict exact‑classification problem, reflecting the ambiguity, breadth, and partial alignment of realistic educational queries.

\item \textbf{An expert‑guided evaluation framework.} We define and apply a candidate-level evaluation protocol in which reference mappings act as semantic anchors while allowing expert judgment over nearest acceptable alternatives, and adopt ECA‑5 (Exact or Closest Acceptable Match at top‑5) as a practically meaningful measure of mapping utility.

\item \textbf{Empirical evidence bridging structural and expert evaluation.}
Through structural and expert‑guided evaluation, we show that structural metrics (EXA‑5 and structural semantic alignment) and expert usefulness capture distinct aspects of system quality, and that CyBOKClaw performs more strongly as a human‑in‑the‑loop candidate generator than as a strict exact classifier.
\end{itemize}
\section{Related Work}
This work lies at the intersection of taxonomy mapping, semantic retrieval, curriculum alignment, and human‑in‑the‑loop evaluation, with a specific focus on aligning cybersecurity curricula to the CyBOK~\cite{cybok_official}.
While each of these areas has been studied extensively in isolation, their integrated treatment within the context of CyBOK educational mapping remains comparatively underexplored.
\subsection{Taxonomy and ontology mapping}
A substantial body of prior work has examined the problem of mapping terms, entities, or concepts into structured taxonomies and ontologies~\cite{ontology_matching_survey,doan2005semantic}.
In ontology alignment and taxonomy mapping, the primary objective is typically to identify semantic correspondences between labels or structured concepts across knowledge systems, often framed in terms of exact concept alignment, semantic similarity, or hierarchical correspondence~\cite{ontology_matching_survey}.

The CyBOK mapping task differs in an important respect. 
Cybersecurity keywords or phrases (KWoPs) arising in educational contexts are frequently broad, lecture‑title-like, or only approximately aligned with CyBOK terminology.
As a result, the task is often not to retrieve a single exact equivalent concept, but to identify one or more semantically acceptable placements within a structured knowledge framework.
Under these conditions, strict tuple‑level alignment provides a useful but incomplete account of practical system performance, motivating evaluation approaches that account for expert‑acceptable semantic proximity rather than exact correspondence alone.

\subsection{Semantic retrieval and top-$k$ candidate generation}
This work is also closely related to research on semantic retrieval and top‑$k$ candidate ranking~\cite{manning2008ir,carpineto2012query,yates2011ir}.
In retrieval‑oriented systems, the objective is often not to produce a single exact answer, but to return a ranked set of plausible candidates that support downstream interpretation or decision‑making.
Techniques such as query expansion, heuristic ranking, semantic enrichment, and top‑$k$ candidate generation are well‑established approaches for improving retrieval performance when direct lexical overlap between queries and targets is limited~\cite{carpineto2012query}.

This perspective is particularly relevant to CyBOK mapping.
Rather than framing the task as one-shot classification, \texttt{CyBOKClaw} is designed to surface a ranked set of plausible CyBOK candidates that narrow the semantic search space for expert review.
In this respect, the framework functions as a semantic retrieval assistant rather than as a rigid closed-set classifier.
This framing matters because many educational queries correspond to topic families, partial concepts, or approximate semantic regions within CyBOK rather than to a single unambiguously correct target row.
\subsection{Human-in-the-loop and expert-supported systems}
This work is also closely related to research on human-in-the-loop and expert-supported decision systems~\cite{horvitz1999hil,voorhees2001ireval}.
In such settings, system quality is not determined solely by exact agreement with a gold-standard label, but by the extent to which the system presents users with sufficiently strong and informative options to support efficient expert judgment.
This distinction is particularly important in domains characterized by ambiguity, partial semantic overlap, or context-dependent interpretation.

The evaluation design adopted in this paper is motivated by this perspective.
Rather than relying exclusively on exact structural matching, we employ candidate-level expert judgment to assess whether the returned top-$k$ set contains an exact or closest acceptable semantic mapping. 
This approach aligns evaluation more closely with expert-facing decision support than with conventional single-label accuracy reporting.
In this sense, the paper contributes a CyBOK-specific application of task-aligned, expert-guided evaluation for human-in-the-loop taxonomy mapping and curriculum alignment.
\subsection{Curriculum alignment and educational knowledge mapping}
There is also relevant prior work in educational knowledge organization and curriculum alignment, where structured frameworks are used to organize, compare, and assess teaching materials and programmes~\cite{harden2001mapping}.
In these settings, mapping educational content to formal frameworks often requires interpretive judgment rather than direct lexical correspondence.
Syllabus topics, module titles, and assessment materials frequently express broad pedagogical themes or abstractions that do not align cleanly with a single entry in a formal taxonomy.

CyBOK mapping exhibits this same characteristic.
For example, lecture titles such as \emph{``Security Overview''} or \emph{``Security Challenges to IoT and Organizations''} typically correspond to families of concepts rather than to a single exact CyBOK row.
As a result, educational knowledge mapping is particularly well suited to retrieval‑oriented and expert‑guided approaches that prioritize semantically acceptable placements over rigid exact matching.
\subsection{Cybersecurity knowledge organization and CyBOK context}
Within the cybersecurity domain, CyBOK provides a comprehensive and structured reference model for organizing cybersecurity knowledge~\cite{cybok_official}.
CyBOK is increasingly used to support curriculum design, programme benchmarking, and educational alignment across academic and professional training contexts~\cite{cyber_education_framework}. 
Despite this growing importance, relatively limited work has examined automated or semi-automated semantic mapping directly into CyBOK, particularly from the perspective of expert-guided top-$k$ retrieval.

This gap defines a useful application niche for the present study.
The contribution is not a claim of broad novelty in taxonomy mapping or semantic retrieval in general; rather, it lies in applying established retrieval and human-in-the-loop principles to the CyBOK-specific educational mapping problem, and in evaluating this setting using expert-guided top-$k$ usefulness metrics alongside complementary structural retrieval measures.
\subsection{Summary}
Taken together, the related literature suggests that CyBOK mapping should not be treated solely as an exact taxonomy classification problem.
Instead, it is more appropriately understood as a semantic retrieval and expert-support task situated at the intersection of ontology alignment, curriculum mapping, and human-in-the-loop evaluation.
The present work contributes to this space by combining a practical CyBOK-specific retrieval framework with an expert-guided evaluation methodology that explicitly captures the distinction between structural exactness and practical semantic usefulness.
\section{System Design and Implementation}
\subsection{System overview}
\begin{figure*}
  \centering
  \includesvg[width=0.85\textwidth]{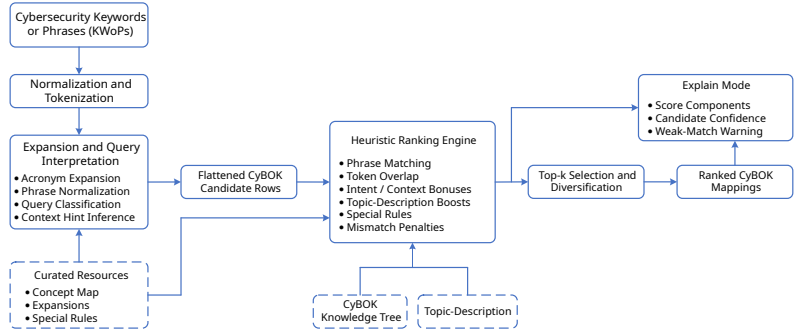}
  \caption{Overview of the \texttt{CyBOKClaw}.
  A cybersecurity keyword or phrase is first normalized and semantically enriched, then ranked against a flattened CyBOK \emph{Knowledge Tree} using a combination of heuristic lexical, semantic, and curated signals.
  The ranked results are subsequently filtered using diversification strategies to produce a top-$k$ candidate list.
  An optional \emph{explain} mode provides score components and confidence labels to support expert review and decision-making.
  }
  \label{fig:cybokclaw}
  \end{figure*}

\texttt{CyBOKClaw}, illustrated in Figure~\ref{fig:cybokclaw}, is a practical semantic retrieval framework designed to map cybersecurity KWoPs to plausible CyBOK candidates. Given an input query, the framework returns a ranked list of \emph{Knowledge Area}, \emph{Topic}, and \emph{Indicative Material} entries from the CyBOK \emph{Knowledge Tree}.
It is intended for human-in-the-loop use, where the highest-ranked candidates support expert judgment and decision-making rather than serving as one-shot final predictions.

The implementation adopts a lightweight yet structured ranking architecture. Its core components include query normalization, curated term expansion, prompt-intent inference, lexical phrase matching, token-overlap scoring, concept-level boosting, topic-description semantic enrichment, special-case ranking rules, and domain-mismatch penalties. 
Final ranking is produced by combining these signals into a single heuristic score, followed by diversification of the returned candidate set. 
The weighting scheme is curated and iteratively refined for the CyBOK setting, reflecting a deliberate trade-off among interpretability, low execution cost, and tunability, while remaining aligned with candidate generation for top-$k$ retrieval rather than strict exact classification~\cite{manning2008ir,yates2011ir}.
\subsection{Data representation}
The core CyBOK \emph{Knowledge Tree}~\cite{cybok_knowledge_tree} is stored as a structured JSON resource, in which each \emph{Knowledge Area (KA)} contains one or more \emph{Topics}, and each Topic contains one or more \emph{Indicative Material (IM)} entries.
At runtime, this hierarchical structure is flattened into candidate rows defined by the combination of \emph{KA}, \emph{Topic}, and \emph{IM} fields, which together form the search space over which ranking is applied.

In addition to the core CyBOK \emph{Knowledge Tree}, the framework draws on several curated supporting resources, including a concept map for phrase-to-row semantic boosts, topic descriptions that enrich rows with aliases and curriculum phrases, curated term expansions for normalizing shorthand variants, and special rules for terms that require domain-sensitive interpretation.
\subsection{Query normalization and expansion}
The first stage of the pipeline normalizes the input query by lowercasing, collapsing whitespace, and tokenizing the text while removing stopwords. 
This produces a cleaned lexical representation suitable for ranking.

The framework then applies term expansion through acronym expansion, phrase normalization, domain-specific term broadening, and concept enrichment for known cybersecurity terminology. This expansion layer is particularly important in cybersecurity, where many queries use abbreviated or cybersecurity-specific forms rather than the exact wording found in CyBOK.
The use of controlled expansion in this way is consistent with established retrieval practice in query expansion and semantic search~\cite{carpineto2012query}.
\subsection{Query classification and context inference}
Because many queries are broad, pedagogically framed, or context-dependent, the framework incorporates a lightweight query-classification stage prior to ranking.
This stage characterizes whether a query is best interpreted as a direct term lookup, a conceptual definition, a broad overview prompt, a risk or governance-oriented query, a cryptography or protocol-oriented query, an AI-for-security query, or another recognizable category.
It also derives context cues such as privacy, network security, cryptography, software security, human factors, risk management, storage security, or formal methods, which are subsequently used to bias ranking toward semantically appropriate regions of CyBOK.
\subsection{Core ranking model}
The ranking model integrates several lightweight scoring components rather than relying on a single retrieval signal.
It accounts for direct and partial phrase matches between the query and candidate entries, employs token-overlap scoring to capture lexical relatedness when exact matches are absent, and applies intent and context-based bonuses to promote semantically appropriate regions of CyBOK.
In addition, curated concept-map boosts and topic-description enrichment reinforce established high-value correspondences and improve robustness when handling broad and pedagogically framed queries. 
Domain-sensitive special-case rules further shape ranking behaviour by promoting salient CyBOK entry families while penalizing alternatives that are lexically plausible but semantically inappropriate, especially in areas such as software supply chain concepts, protocol families, PKI and certificate management, link-layer attacks, and AI-related security queries.
\subsection{Topic-family shaping and mismatch control}
A key design feature of the framework is the use of domain-mismatch penalties to suppress semantically implausible CyBOK entries that might otherwise rank highly due to superficial lexical overlap.
Drift toward generic fallbacks such as \emph{webification}, server-side misconfiguration entries, physical attacks, or loosely related AI-centric topics is mitigated by penalizing topic families that are inconsistent with the inferred domain of the query.
This mechanism proved particularly important in improving validation performance on protocol-oriented, PKI-related, and infrastructure-heavy queries where lexical ambiguity is common and structural cues play a critical role in effective ranking.
\subsection{Top-$k$ diversification}
After scoring, the framework orders candidate entries in descending score and applies a lightweight diversification step to prevent the final top-$k$ list from being dominated by near-duplicate rows drawn from the same topic grouping.
This stage enhances the practical value of the returned candidate set for expert review by balancing relevance with controlled topical diversity, in line with established practices in top-$k$ retrieval and ranked candidate presentation~\cite{yates2011ir,jarvelin2002topk}.
\subsection{Explain mode and confidence reporting}
The framework incorporates an explain mode that reports each candidate rank, numerical score, constituent score components, and an overall confidence label expressed as \textbf{strong}, \textbf{moderate}, or \textbf{weak}.
These labels are derived directly from the final heuristic score. 
In addition, explain mode provides query-level annotations indicating whether the top-ranked candidates exhibit strong heuristic alignment, only moderate or topic-level alignment rather than row-level correspondence, or no strong CyBOK match.
This functionality improves transparency and reduces the risk that weak heuristic results are misinterpreted as high-confidence exact mappings.
In this respect, the design aligns with broader human-in-the-loop and mixed-initiative principles, in which explanatory support is intended to inform and augment expert judgment rather than replacing it~\cite{horvitz1999hil}.
\subsection{Implementation philosophy}
The framework deliberately adopts a curated heuristic architecture because it offers several advantages in the present context, including interpretability, tunability, computational efficiency, and controllability. 
In practical terms, ranking behaviour can be inspected and explained, specific concept families can be iteratively refined, the system can be executed efficiently without heavy external dependencies, and domain-specific corrections can be introduced directly through curated resources. 
The principal trade-off is that overall performance depends on the breadth and quality of the curation layer.
As demonstrated in the evaluation, targeted refinement of expansions, concept maps, and topic descriptions can substantially improve generalization, but this also entails an ongoing maintenance burden.
\subsection{Operational Scope}
At its present stage, \texttt{CyBOKClaw} operates as a strong top-$k$ CyBOK candidate generator supporting single-keyword and curriculum-style query mapping, explain mode, hierarchy browsing, batch mapping, confidence reporting, and curated concept support for supply-chain, protocol, PKI, and infrastructure queries.
The framework also provides integrated benchmarking and evaluation workflows across development and validation datasets.
\texttt{CyBOKClaw} is therefore best understood as an expert-support semantic retrieval system whose primary objective is to deliver useful candidate mappings efficiently and transparently, rather than to guarantee fully autonomous exact classification in all cases. 
As its name suggests, \texttt{CyBOKClaw} is implemented as an \emph{Agent Skill} within the OpenClaw~\cite{openclaw} environment, but the framework can also be operated independently through a standalone Python implementation.
The authors will be able to demonstrate the capabilities of \texttt{CyBOKClaw} if the paper is accepted for presentation at the Advances in Teaching and Learning for Cyber Security Education 2026 (ATLCS26) conference.
\section{Evaluation Methodology}
\subsection{Overview}
The evaluation was designed to assess how effectively \texttt{CyBOKClaw} retrieves useful CyBOK mappings for cybersecurity KWoPs. 
Because many input queries are broad, ambiguous, or only approximately aligned with CyBOK terminology, the framework was evaluated using a combination of \emph{structural retrieval metrics} and \emph{expert-guided semantic judgment}.
In such cases, the system may fail to retrieve the exact reference tuple while nevertheless returning a candidate that would be acceptable in expert practice.
Accordingly, the evaluation is organized around two complementary components: structural evaluation, which measures exact and semantically aligned retrieval against the human reference mappings, and expert-guided semantic evaluation, which assesses whether the returned top-$k$ candidates contain at least one mapping that an expert reviewer would judge as acceptable.
\subsection{Evaluation corpora}
The experiments employ two sets of datasets: a development set, summarized in Table~\ref{tab:dev-datasets}, comprising eight cybersecurity curriculum-derived datasets, and a validation set, summarized in Table~\ref{tab:val-datasets}, comprising three additional datasets used to assess generalization beyond the development material. 
Each dataset contains a collection of KWoPs paired with a human-produced CyBOK reference mapping.
The reference mapping specifies a \emph{Knowledge Area (KA)}, \emph{Topic}, and \emph{Indicative Material (IM)}, and serves as the semantic anchor for both structural and expert-guided evaluation. 
Reference mappings for five cybersecurity modules: Security 101, Introduction to Cybersecurity, Applied Cyber Security, Security Principles, and Information Security Management (COMM037) were produced by the CyBOK project team.
Reference mappings for the remaining three development datasets, as well as the validation datasets, were produced by the author(s).
\begin{table}[h]
\centering
\caption{Development datasets with eight cybersecurity modules.}
\label{tab:dev-datasets}
\begin{tabular}{>{\raggedright\arraybackslash}p{5.50cm}|>{\raggedright\arraybackslash}p{4.60cm}|>{\centering\arraybackslash}p{1.70cm}}
\toprule
\textbf{Module Name} & \textbf{Institution} & \texttt{\#} KWoPs \\
\midrule
Security 101~\cite{cybok2020bristol}                             & University of Bristol, UK                  & 30 \\
Introduction to Cybersecurity~\cite{cybok2020harvard}            & Harvard University, USA                    & 25 \\
Applied Cyber Security~\cite{cybok2020mit}                       & Massachusetts Institute of Technology, USA & 40 \\
Security Principles~\cite{cybok2020oxford}                       & University of Oxford, UK                   & 49 \\
Information Security Management (COMM037)~\cite{cybok2020surrey} & University of Surrey, UK                   & 23 \\
Fundamentals of Networks and Security                            & University of Derby                        & 39 \\
Ethical Hacking                                                  & University of Derby                        & 40 \\
Security Management                                              & University of Derby                        & 25 \\
\bottomrule
\end{tabular}
\end{table}
\begin{table}[h]
\centering
\caption{Validation datasets with three cybersecurity modules.}
\label{tab:val-datasets}
\begin{tabular}{>{\raggedright\arraybackslash}p{5.50cm}|>{\raggedright\arraybackslash}p{4.60cm}|>{\centering\arraybackslash}p{1.70cm}}
\toprule
\textbf{Module Name} & \textbf{Institution} & \texttt{\#} KWoPs \\
\midrule
Communication and Security Protocols     & University of Derby & 47 \\
Security Architectures and Future Trends & University of Derby & 40 \\
Artificial Intelligence for Security     & University of Derby & 13 \\
\bottomrule
\end{tabular}
\end{table}
\subsection{Structural evaluation}
\subsubsection{EXA-5}
For each query, the framework returns the top-5 ranked CyBOK candidates.
\textbf{EXA-5} measures whether the human reference mapping appears anywhere within this top-5 set.
This metric provides a strict assessment of retrieval precision based on exact tuple-level matching.
\subsubsection{Structural semantic alignment}
Because exact tuple matching is often too restrictive for semantically broad or approximate queries, we additionally compute \textbf{structural semantic alignment}, which assesses whether the returned candidate set is structurally and semantically aligned with the reference mapping. 
This metric is less strict than exact matching but remains fully algorithmic rather than expert-judged~\cite{voorhees2005trec,jarvelin2002topk,sakai2007exact}.
\subsection{Expert-guided semantic evaluation}
\subsubsection{Candidate-level judgment}
For each query, the top-5 returned candidates are evaluated with respect to both the semantic intent of the original query and the human reference mapping, which serves as a semantic anchor. 
Candidate-level review was conducted using a single-expert judgment workflow guided by written category definitions. 
The expert reviewer was familiar with cybersecurity curriculum mapping and CyBOK structure, and applied a predefined rubric to assign one of four labels to each candidate: \textbf{Exact}, \textbf{Closest Acceptable}, \textbf{Relevant but not Closest}, or \textbf{Not Acceptable}. 
The current study employs a single-expert protocol. 
Consequently, the expert-guided results should be interpreted as a task-aligned utility assessment rather than a consensus semantic gold standard.
\subsubsection{Handling partially specified reference mappings}
Some human reference mappings include \texttt{***} in the Topic or Indicative Material fields. These markings are treated as intentional expert annotations rather than missing data.
If \textbf{IM = \texttt{***}}, the reference is interpreted primarily at \textbf{KA} and \textbf{Topic} level.
If \textbf{Topic = \texttt{***}} and \textbf{IM = \texttt{***}}, the reference is interpreted primarily at the \textbf{KA} level.
If \textbf{KA = \texttt{***}}, the item is handled with caution and treated as potentially out of scope or lacking a direct CyBOK anchor.
This granularity-aware interpretation prevents the evaluation from imposing inappropriate tuple-level exactness where the human reference mapping is deliberately broader in scope.
\subsubsection{ECA-5}
The primary expert-guided usefulness metric employed in this work is \textbf{ECA-5 (Exact or Closest Acceptable Match at top-5)}. 
A query is counted as an ECA-5 success if at least one of the top-5 returned candidates is labeled either \textbf{Exact} or \textbf{Closest Acceptable}. ECA-5 is well suited for \texttt{CyBOKClaw} because the framework is designed to support expert review through top-$k$ candidate generation rather than to function as a single-label exact matcher~\cite{voorhees2005trec,jarvelin2002topk}.
This expert-guided evaluation protocol should therefore be interpreted as a task-aligned assessment of practical utility rather than as a claim of universal semantic ground truth.
\subsection{Rationale for combined structural and expert‑guided evaluation}
The use of both structural and expert-guided evaluation provides a more comprehensive account of framework quality.
EXA-5 measures strict tuple-level retrieval precision, structural semantic alignment measures algorithmic proximity within the CyBOK structure, and ECA-5 assesses practical, expert-acceptable usefulness.
While structural semantic alignment reflects how closely retrieved candidates align with the reference mapping in a formal sense, ECA-5 captures whether the returned candidate set would be genuinely usable by an expert reviewer.
Taken together, these complementary metrics distinguish between performance derived from exact matching and performance derived from effectively surfacing semantically useful candidate mappings.
\subsection{Iterative workflow}
The evaluation and refinement process followed an iterative workflow centred on the development split. 
The framework was initially analysed on the development datasets to identify systematic failure patterns, after which the curation layer and ranking rules were refined and structural benchmarks were re-executed. 
The validation split was then used as a held-out assessment of how well the resulting framework generalized beyond the development material. 
Expert annotations and corresponding ECA-5 computation were subsequently performed over the benchmark outputs.
This workflow supported both progressive framework improvement and a more detailed understanding of where structural metrics and expert-guided usefulness diverge.
\section{Results}
\subsection{Overview}
We evaluate \texttt{CyBOKClaw} using a combination of structural retrieval metrics and expert‑guided semantic evaluation.
In this paper, \textbf{EXA‑5} denotes strict recovery of the human reference mapping within the top‑5 returned candidates, \textbf{Structural Semantic} alignment measures algorithmic proximity to the reference mapping within the CyBOK structure, and \textbf{ECA‑5} denotes expert‑guided top‑5 success, defined as the presence of at least one candidate judged as \emph{Exact} or \emph{Closest Acceptable}.
Taken together, these metrics distinguish strict retrieval performance from practical candidate‑set usefulness.

A simple divergence case helps illustrate this distinction.
A query may fail EXA‑5 because the exact reference row does not appear in the top‑5 results, while still achieving a strong structural semantic alignment score if the retrieved candidates occupy the correct region of the CyBOK \emph{Knowledge Tree}.
The same query may nevertheless count as an ECA‑5 success if an expert judges one of those candidates to be the nearest acceptable practical mapping.
Across both the development and validation datasets, exact tuple‑level retrieval consistently underperforms expert‑guided usefulness.
This pattern indicates that the framework is often more effective as a ranked candidate generator supporting expert decision‑making than strict exact‑match reporting alone would suggest.
\subsection{Overall Evaluation Results}
In the evaluation, both structural and expert‑guided metrics are computed over exactly the same set of queries.
On the development split, \texttt{CyBOKClaw} achieves 64.73\% EXA-5, 84.18\% structural semantic alignment, and 91.88\% ECA-5 across 271 queries.
On the validation split, it achieves 81.19\% EXA-5, 93.32\% structural semantic alignment, and 98.00\% ECA-5 across 100 queries.
\begin{table}[h]
\centering
\caption{Overall evaluation results on development and validation splits.}
\label{tab:overall-results}
\begin{tabular}{>
{\raggedright\arraybackslash}p{5.00cm}>{\centering\arraybackslash}p{1.30cm}>{\raggedleft\arraybackslash}p{1.70cm}>{\raggedleft\arraybackslash}p{2.00cm}>{\raggedleft\arraybackslash}p{1.70cm}
}
\toprule
\textbf{Split} & \textbf{KWoPs} & \textbf{EXA-5} & \textbf{Structural Semantic} & \textbf{ECA-5} \\
\midrule
Development & 271 & 64.73\% & 84.18\% & 91.88\% \\
Validation & 100 & 81.19\% & 93.32\% & 98.00\% \\
\bottomrule
\end{tabular}
\end{table}

These results illustrate a clear pattern: exact row-level retrieval is strong, structural semantic alignment is stronger, and expert-guided ECA-5 is strongest.
On the development split, ECA-5 exceeds EXA-5 by 27.15\%; on the validation split, the corresponding gap is 16.81\%.
In practical terms, the framework typically retrieves at least one top-5 candidate that an expert would accept as an appropriate CyBOK mapping, even in cases where the precise reference row is not returned.
\subsection{Dataset‑Level Evaluation Results}
Table~\ref{tab:dataset-level-results} reports the dataset-level evaluation results.
Several broad or curriculum-oriented datasets, including Harvard, Oxford, and Surrey, remain challenging under strict exact-match evaluation, yet still achieve strong ECA-5 scores.
This pattern is consistent with the intended role of \texttt{CyBOKClaw} as a human-in-the-loop candidate generator rather than a strict exact-matching system.
\begin{table*}[t]
\centering
\caption{Dataset-level evaluation results.}
\label{tab:dataset-level-results}
\footnotesize
\begin{tabular}{>{\raggedright\arraybackslash}p{5.90cm}>{\centering\arraybackslash}p{1.30cm}>{\raggedleft\arraybackslash}p{1.35cm}>{\raggedleft\arraybackslash}p{1.80cm}>{\raggedleft\arraybackslash}p{1.35cm}}
\toprule
\textbf{Dataset} & \textbf{KWoPs} & \textbf{EXA-5} & \textbf{Structural Semantic} & \textbf{ECA-5} \\
\midrule
Bristol                                  & 30 & 60.00\% & 78.33\% &  90.00\% \\
Ethical Hacking                          & 40 & 76.19\% & 86.90\% &  95.00\% \\
Fundamentals of Networks and Security    & 39 & 78.05\% & 84.76\% &  92.31\% \\
Harvard                                  & 25 & 48.00\% & 82.00\% &  96.00\% \\
MIT                                      & 40 & 60.00\% & 87.50\% &  87.50\% \\
Oxford                                   & 49 & 59.18\% & 83.16\% &  91.84\% \\
Security Management                      & 25 & 76.00\% & 88.00\% &  92.00\% \\
Surrey                                   & 23 & 52.17\% & 80.43\% &  91.30\% \\
Artificial Intelligence for Security     & 13 & 53.85\% & 86.54\% &  92.31\% \\
Communication and Security Protocols     & 47 & 82.98\% & 92.55\% &  97.87\% \\
Security Architectures and Future Trends & 40 & 87.80\% & 96.34\% & 100.00\% \\
\bottomrule
\end{tabular}
\end{table*}

Performance on the validation split is particularly strong. 
All validation datasets exceed 92\% ECA-5, with one achieving 100\%. 
Within the development split, Harvard provides the clearest example of the gap between exact retrieval and expert-guided usefulness: although EXA-5 is 48.00\%, ECA-5 reaches 96.00\%. 
Oxford and Surrey show a similar pattern, indicating that many difficult curriculum-style queries remain practically usable for expert review even when the exact reference row is not retrieved. 
The MIT dataset represents an instructive boundary case in which structural semantic alignment and ECA-5 coincide at 87.50\%, indicating closer agreement between algorithmic proximity and expert-accepted usefulness.
\subsection{Interpretation of structural–expert metric gaps}
A substantial gap between EXA‑5 and ECA‑5 is observed on both the development and validation splits, with the effect particularly pronounced on the development dataset.
This divergence is not merely a scoring artefact; rather, it indicates that many framework outputs are semantically useful for expert review even when the exact reference tuple does not appear within the top‑5 results.
Structural semantic alignment captures part of this phenomenon algorithmically by reflecting proximity within the CyBOK \emph{Knowledge Tree}, whereas ECA‑5 captures it more directly through expert‑guided judgments of acceptability.

From the perspective of intended use, this gap constitutes a central finding.
\texttt{CyBOKClaw} is most effective as an expert‑support retrieval system that reduces the search space, surfaces semantically plausible CyBOK candidates, and lowers the cognitive and manual effort required for final expert mapping decisions.
\subsection{Worked example}
A representative success case is the Oxford query \emph{``Secure sockets layer''}.
In the expert evaluation, the top‑ranked candidate returned by \texttt{CyBOKClaw} is \emph{Network Security $\rightarrow$ Network Protocols and Their Security $\rightarrow$ Security at the Transport Layer}.
This candidate was judged \emph{Closest Acceptable} rather than \emph{Exact}.
The example is instructive because it highlights the distinction between exact row‑level recovery and practical expert usefulness: even when the preferred human reference is phrased differently or anchored at a slightly different level of granularity, the framework surfaces a clearly defensible CyBOK placement within the top‑5 results.

A contrasting set of harder cases involves broad curriculum‑style prompts such as \emph{``Security Overview''} or architecture‑heavy inputs that span multiple CyBOK regions.
In such cases, the framework may correctly identify the relevant topic family or conceptual area without selecting a single best exact row.
These cases disproportionately affect EXA‑5 while largely preserving ECA‑5, helping to explain why expert‑guided usefulness remains high even when strict tuple‑level recovery is imperfect.
\section{Discussion}
The experimental results indicate that \texttt{CyBOKClaw} is substantially stronger as a semantic candidate‑generation framework than a strict exact‑match perspective alone would suggest.
Across both the development and validation datasets, ECA‑5 consistently exceeds EXA‑5, indicating that the framework frequently returns useful candidate mappings even when it fails to recover the precise reference row.

This gap is particularly informative on the development dataset.
Although the framework does not always identify the single preferred exact row, it reliably retrieves candidates from the correct semantic region and surfaces at least one mapping that an expert would accept in practice.
For human‑in‑the‑loop CyBOK mapping, this behaviour reflects a more meaningful notion of system usefulness than exact tuple equality alone.

The dataset-level results further sharpen this interpretation.
Broad, conceptual, or curriculum‑oriented datasets such as Harvard, Oxford, and Surrey remain comparatively challenging under strict exact‑match evaluation, yet achieve strong ECA‑5 scores under expert review.
This suggests that the principal residual difficulty often lies not in selecting the appropriate semantic region of CyBOK, but in identifying the single best row within an already relevant area.
In contrast, the validation datasets indicate that when queries are more locally grounded or structurally constrained, the framework can perform strongly across all three metrics simultaneously.

The comparison between structural and expert‑guided evaluation also has important methodological implications. 
Exact structural matching captures canonical precision, while structural semantic alignment measures algorithmic proximity within the taxonomy.
ECA‑5, by contrast, evaluates whether the returned candidate set is practically usable under an explicit expert review protocol.
In the CyBOK context, where queries are frequently broad, educational in nature, and only partially aligned with formal taxonomy phrasing, this distinction is consequential. 
The results therefore support expert‑guided top‑$k$ evaluation as a task‑aligned complement to exact structural metrics~\cite{voorhees2005trec,jarvelin2002topk,sakai2007exact}.

At the same time, the framework does not perform uniformly across all query types.
Exact row‑level precision remains weaker than top‑$k$ usefulness, and some broad, architecture‑heavy, or protocol‑sensitive queries continue to yield approximate rather than fully exact outputs.
Accordingly, \texttt{CyBOKClaw} is best understood as an expert‑support retrieval system whose primary value lies in narrowing the semantic search space and improving candidate quality to support efficient and informed human mapping decisions.

\subsection{Limitations}
This work has several limitations. 
Firstly, the framework is based on a curated heuristic retrieval architecture, and its performance therefore depends materially on the breadth and quality of constructed resources, including term expansions, concept maps, topic descriptions, and ranking rules. 
Reported performance thus reflects not only retrieval design choices but also curation effort, and portability to other taxonomies or domains should not be assumed without additional adaptation. 
Second, while expert-guided evaluation provides a more faithful account of practical usefulness than exact matching alone, expert judgment inherently introduces an element of interpretive subjectivity. 
The evaluation results reported in this paper employ a single-expert workflow with iterative consistency checks, rather than a multi-annotator adjudication design. 
As a result, the reported ECA-5 scores should be interpreted as a task-aligned expert utility measure rather than as a full inter-rater reliability study. 
Third, the current paper does not include direct baseline comparisons against simpler lexical or embedding-based retrieval systems, so the results should be interpreted primarily as evidence that the proposed framework is useful in absolute and task-aligned terms rather than as a definitive comparative benchmark. 
Fourth, the framework is consistently more effective at identifying the correct semantic region within CyBOK than at selecting the single best exact row, particularly for broad, architecture-heavy, or protocol-sensitive queries. 
Finally, while the reported results reflect the scope and characteristics of the current benchmark datasets, they do not encompass the full diversity of cybersecurity terminology, educational styles, or curriculum formulations encountered in practice.
\subsection{Future Work}
A natural extension of \texttt{CyBOKClaw} is a hybrid, Large Language Model (LLM)-assisted architecture in which the current heuristic framework generates a bounded top-$k$ set of CyBOK candidates, followed by second-stage semantic reranking using a LLM. 
Such a design could preserve the efficiency, interpretability, and controllability of the existing architecture while improving disambiguation for broad, infrastructure-heavy, acronym-dense, or protocol-sensitive queries. 
This extension is particularly well aligned with the framework’s primary residual limitation: selection of the single best exact row within an already correct topic family.

More broadly, future work should advance along four complementary directions. 
First, exact row-level precision within correctly identified topic families should be improved, particularly for protocol and infrastructure-oriented concepts. 
Second, the curation layer should be expanded and systematically regularized to reduce reliance on ad hoc coverage and incremental tuning. Third, direct baseline comparisons against simpler lexical and semantic retrieval methods should be added to clarify the empirical contribution of the current architecture. 
Fourth, a direct empirical comparison between the current heuristic framework and a hybrid heuristic--LLM variant would help clarify trade-offs among semantic flexibility, interpretability, reproducibility, and computational cost.
\section{Error Analysis}
The remaining errors in \texttt{CyBOKClaw} are concentrated in a small number of recurring patterns that illuminate both current limitations and opportunities for further refinement.
The most common failure mode arises when the framework retrieves candidates from the correct semantic region of CyBOK but does not identify the single best exact row, leading to Closest Acceptable rather than Exact judgments and accounting for much of the residual gap between EXA‑5 and ECA‑5.

Broad, curriculum-style queries continue to pose challenges because they naturally align with topic families rather than individual rows, disproportionately penalizing strict exact-match metrics despite yielding semantically appropriate candidate sets. 
Infrastructure and architecture-heavy queries such as those involving large-scale networks, IoT deployments, cloud access models, or enterprise architectures remain more error-prone because of their cross-domain nature and weak localization to individual CyBOK entries.
Protocol and standards-oriented queries similarly expose difficulties in layer and context disambiguation, particularly where cryptographic, operational, and infrastructure meanings intersect. 
While curated semantic resources have proven effective in mitigating many of these issues, performance remains sensitive to coverage gaps in the curation layer.

Collectively, these patterns suggest that future improvements should focus on enhancing exact row selection within already correct topic families, strengthening disambiguation for cross‑cutting and protocol‑sensitive concepts, and systematically expanding curated semantic coverage reinforcing \texttt{CyBOKClaw}’s role as a progressively improving expert‑support retrieval system rather than a fully autonomous classifier.
\section{Conclusion}
This paper presented \texttt{CyBOKClaw}, a practical semantic retrieval framework for aligning cybersecurity KWoPs with the CyBOK. 
The framework is explicitly designed as a human-in-the-loop top-$k$ mapping assistant, reflecting the ambiguity, breadth, and pedagogical framing of realistic educational queries.
The experimental results indicate that structural and expert-guided metrics capture distinct and complementary aspects of framework quality. 
On the development datasets, \texttt{CyBOKClaw} achieved 64.73\% EXA-5, 84.18\% structural semantic alignment, and 91.88\% ECA-5, while on the validation datasets it achieved 81.19\% EXA-5, 93.32\% structural semantic alignment, and 98.00\% ECA-5. 
Taken together, these results suggest that \texttt{CyBOKClaw} is substantially more effective as an expert-support candidate-generation system than would be suggested by exact-match retrieval performance alone.

The primary contribution of this study is a CyBOK-specific formulation of curriculum-oriented mapping as a human-in-the-loop top-$k$ retrieval problem, together with a combined structural and expert-guided evaluation methodology that more directly reflects expert-facing utility. 
The findings show that expert-acceptable top-$k$ retrieval provides a more practical and informative account of CyBOK mapping quality than strict exact structural matching in isolation. 
Future work should focus on improving exact row-level precision within already correct topic families, extending semantic support for protocol and infrastructure-heavy concepts, and empirically evaluating bounded hybrid reranking architectures that combine heuristic retrieval with semantic disambiguation. 
Overall, the results support viewing \texttt{CyBOKClaw} as a practical CyBOK mapping assistant and reinforce the value of task-aligned, expert-guided evaluation for understanding and designing expert-facing taxonomy mapping systems.
\bibliographystyle{splncs04}
\bibliography{references}
\appendix
\section{CyBOK Mapping for Certification}
In the United Kingdom (UK), the National Cyber Security Centre (NCSC) operates a certification scheme for cybersecurity degree apprenticeships, undergraduate programmes, integrated master’s degrees, and postgraduate master’s programmes.
As part of this certification process, institutions are required to demonstrate appropriate coverage of the CyBOK across their curricula.
Coverage is evidenced through the allocation of academic credits, where one credit corresponds to a notional ten hours of student learning.
Applying institutions must therefore map taught modules, learning outcomes, and assessment content to CyBOK Knowledge Areas, Topics, and Indicative Material, and aggregate associated credits to demonstrate compliance with the NCSC coverage requirements.

This mapping process is non-trivial in practice. 
Curriculum materials are typically expressed using course-specific terminology, broad module titles, or pedagogical abstractions that do not align cleanly with CyBOK terminology. 
\texttt{CyBOKClaw} directly supports this certification workflow by functioning as an expert-support semantic retrieval tool that maps cybersecurity keywords and curriculum-style queries to plausible CyBOK candidates. 
By providing ranked top-$k$ CyBOK mappings, explainable scores, and confidence indicators, the framework assists academic staff and programme designers in efficiently identifying appropriate CyBOK placements, narrowing the search space for expert judgment, and supporting transparent, auditable CyBOK credit allocation for NCSC certification submissions.
\begin{figure}[t]
  \centering
  \includesvg[width=0.85\textwidth]{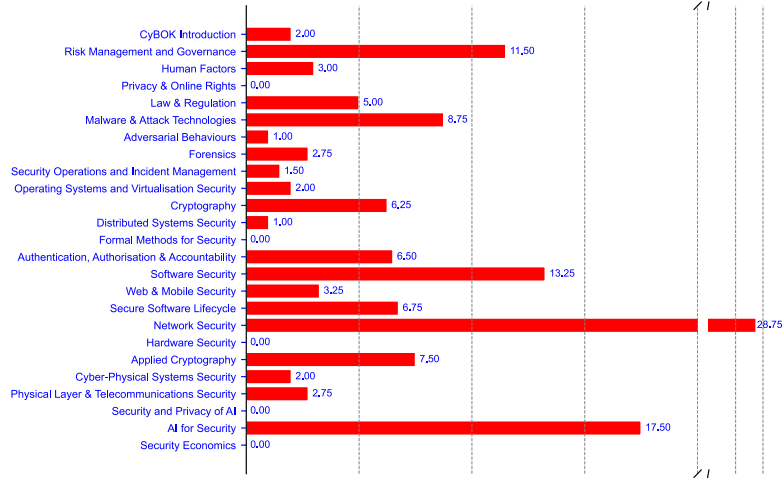}
  \caption{Example of CyBOK mapping and credit allocation. Module‑level mappings and associated credits are aggregated by CyBOK Knowledge Area to produce a curriculum‑level coverage profile.}
  \label{fig:uod-ug-cybok-barh}
\end{figure}

Figure~\ref{fig:uod-ug-cybok-barh} presents the CyBOK mappings and associated credit allocation for six cybersecurity modules from the University of Derby comprising three modules from the development set and three from the validation set.
Figure~\ref{fig:uod-ug-cybok-barh} also includes an additional 13 credits drawn from three further modules that contain partial coverage of cybersecurity content.

In CyBOK, a Broad Category is a high-level grouping of Knowledge Areas used to organize the overall scope of cybersecurity knowledge into coherent conceptual domains. 
Broad Categories do not constitute an additional formal taxonomy layer such as \emph{Knowledge Area} $\rightarrow$ \emph{Topic} $\rightarrow$ \emph{Indicative Material}; instead, they serve as an organizational and reporting abstraction that supports communication, curriculum design, and certification evidence.

\begin{figure}[t]
  \centering
  \includesvg[width=0.85\textwidth]{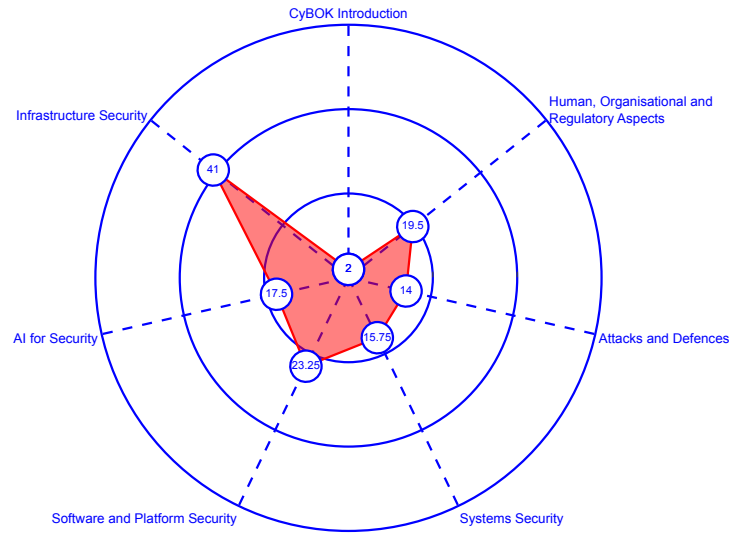}
  \caption{Coverage across CyBOK broad categories. Note that AI for Security is not a CyBOK broad category but a topic guide, and is included here to illustrate the curriculum’s coverage of AI‑related cybersecurity content.}
  \label{fig:uod-ug-cybok-spider}
\end{figure}

Figure~\ref{fig:uod-ug-cybok-spider} presents a spider-map visualization of a cybersecurity curriculum following the CyBOK mapping and credit allocation shown in Figure~\ref{fig:uod-ug-cybok-barh}, with aggregated coverage reported across CyBOK Broad Categories.
%
%
\end{document}